\documentclass[a4paper,12pt]{article}
\usepackage{graphicx}
\usepackage{bm}
\usepackage{amstext,amsmath,amssymb}
\usepackage{mathptmx}
\usepackage[top=3cm,bottom=3cm,left=1.5cm,right=1.5cm,asymmetric]{geometry}
\usepackage{rotating}
\usepackage{hyperref}
\usepackage{cite}
\usepackage{array}
\usepackage{tabularx}
\usepackage{caption}
\begin{document}

\title{Comparative insights into gluon and proton structure through parton distribution functions}
\author{\bf{Akbari Jahan \textsuperscript{1,a} $\mathrm{and}$ Diptimonta Neog \textsuperscript{2}}\\
{} \textsuperscript{1,2}Department of Physics, North Eastern Regional Institute of Science and Technology, \\ Nirjuli-791109, Arunachal Pradesh, India. \\
{} \textsuperscript{a}E-mail: akbari.jahan@gmail.com} 
\date{}
\maketitle

\abstract{Study of parton distribution functions (PDFs) has led to a finer cognisance of the structure of partons in hadrons and the proton structure functions in deep inelastic scattering (DIS). PDFs are instrumental in predicting results for most of the hard-scattering processes measured at the Large Hadron Collider (LHC). However, due to the non-perturbative nature of partons, calculation of parton distributions using perturbative QCD (Quantum Chromodynamics) cannot be done. The analysis of PDFs, therefore, needs a relentless effort. In this paper, we study a comparative approach of three global PDF sets, viz. CT14, MMHT2014 and NNPDF 4.0. Gluon distribution functions and proton structure functions have been evaluated in a wide range of momentum fraction \textit{x} and energy scale \textit{Q}; and plausible observations have been made.
}

\section{Introduction}
\label{section1}
The substructure of hadrons apropos quarks and gluons (partons) are explained in detail by parton distribution functions (PDFs). Partons are the basic degrees of freedom of Quantum Chromodynamics (QCD), the theory of the strong interaction between quarks and mediated by gluons. Precise comprehension of the PDFs has become indispensable with the development of the Large Hadron Collider (LHC) to study the fundamental interactions of nature. This is due to the fact that the cross section calculations of collider experiments employ PDFs to a large extent ~\cite{Lai}. During the proton-proton collision, it is the partons that encounter hard collisions. The interaction of quarks and gluons inside the proton result in momenta distribution that are studied in respect of PDFs ~\cite{Forte, Forte-Watt}. They are best understood by using the global fits to the available data; for example, deep inelastic scattering (DIS) data from HERA (Hadron Electron Ring Accelerator) provide enough information about the wide gluon density at small \textit{x} ~\cite{Devenish}. In other words, PDFs are ascertained by global analyses of data obtained from DIS experiments. The production of heavy quarks in hadron-hadron and lepton-hadron collisions gives a fine test of QCD. The global analysis of parton distributions utilise data of several hard-scattering experiments. Deep inelastic scattering is one such main source of information on parton distributions. While NNPDF (Neural Network PDF) collaboration ~\cite{Forte-Garrido} is a global fit ~\cite{Ball1} based on DIS data ~\cite{Ball2}, MMHT (Martin-Motylinski-Harland-Lang-Thorne) ~\cite{Harland} and CTEQ/CT (Coordinated Theoretical Experimental project on QCD) ~\cite{Hou} collaborations are focussed on determining PDF uncertainties.\\

PDFs employed in hard-scattering processes are solutions of DGLAP (Dokshitzer-Gribov-Lipatov-Altarelli-Parisi) equations that evolve PDFs from a starting scale, say $Q_0$ , to any higher scale. The scale dependence of parton distributions, governed by these integro-differential QCD evolution equations, is initiated by the interactions of partons through processes, viz. $q \rightarrow qg $ (emission of gluons from quarks), $g \rightarrow q \bar q $ (quark-antiquark pair production by gluons) and $g \rightarrow gg $ (emission of gluons from gluons) ~\cite{Gribov, Altarelli, Dokshitzer}. The physics behind this scale dependence of parton distributions is not difficult to comprehend. When a parton, carrying momentum fraction \textit{x}, radiates another parton, its momentum fraction gets reduced and thus the parton density at that value of \textit{x} also decreases. Nevertheless, the parton density at that value of \textit{x} will also acquire larger momentum from other partons with higher momentum fractions emitting partons. There is thus a scenario wherein momentum is gained and reduced alternately. Nearly half of the proton’s momentum is carried by gluons, wherein most of the momentum is present at small \textit{x}. While both up and down quarks overtop at large \textit{x}, gluons are dominant at small \textit{x}. The behaviour of partons at small and large \textit{x} are generally related by momentum sum rule ~\cite{Campbell}.\\

A parton distribution function $f_i(x,Q)$, where \textit{i} represents parton species, depends on two variables: momentum fraction \textit{x} carried by the parton and momentum scale \textit{Q} at which the nucleon is observed. The PDFs are parametrized at a low momentum initial scale $Q_0$, of order $\sim$ 1 GeV, by a standard functional form with variable parameters $(c_0,c_1,c_2,...)$.\\
\begin{equation}
f(x,Q_0)= c_0 \, x^{c_1} \, (1-x)^{c_2}\, A(x)
\end{equation}

Here $A(x)$ is a smooth function with a few more free parameters and it remains finite both when $x \rightarrow 0 $ and $ x \rightarrow 1 $. The \textit{Q} dependence of $f(x,Q)$ is determined by the integro-differential QCD evolution equations ~\cite{Gribov, Altarelli, Dokshitzer}.\\

At asymptotically small values of \textit{x}, PDFs behave as $x^{c_1}$ (when $x \rightarrow 0 $) and are explicated by the Regge theory ~\cite{Regge, Ball-Nocera}. On the other hand, at asymptotically large values of \textit{x}, the PDFs behave as $(1-x)^{c_2}$ (when $ x \rightarrow 1 $) and are elucidated by quark counting rules ~\cite{Brodsky}. The outset of such an asymptotic order depends not only on the PDF flavour but also the global PDF set under consideration.\\

As PDFs are paramount in the explicit study of LHC data, their definite calculation is a must. The LHC data helps in distinguishing different PDF sets and thus assist in improving the accurate determination of PDFs ~\cite{Forte-Watt, Rojo, Ball3, Butterworth, Jimenez}. However, due to the present constraints in the conception of non-perturbative QCD, such a calculation is not feasible from the first principles. Rather, the functional form of PDFs are usually identified from global fit data of hard-scattering experiments ~\cite{Harland, NNPDF, Dulat, Alekhin, ZEUS}. The above equation is, in fact, motivated from theoretical estimation of power-law behaviour of PDFs at asymptotically large and small values of \textit{x}, and is based on non-perturbative QCD approaches ~\cite{Devenish, Roberts}. Several authors and collaborations have recently worked on the global QCD analyses of parton distributions ~\cite{Hou, Ball4, Barry1, Bailey, Han, Barry2}.\\

The aim of this paper is to briefly explore the significance of PDFs, that are relevant for understanding the hard processes of QCD. Comparisons of PDFs from the global fitting collaborations have been discussed in relation to the plots obtained using APFEL, a PDF Evolution Library ~\cite{Bertone, Carrazza}. The layout of the paper is as follows. In section ~\ref{section2}, the methodology of the proposed study has been clearly stated. Section ~\ref{section3} discusses the significant observations and results that have been made based on the PDF plots. Conclusion and future outlook are given in section ~\ref{section4}.

\section{Methodology}
\label{section2}
In this paper, APFEL has been used as the PDF tool. It is a computer library associated with the solutions of DGLAP evolution equations. The graphical depiction of PDFs are well manifested for various data/inputs. As mentioned earlier, PDFs cannot be computed from the first principles; rather their evaluation is carried out by making proper predictions of hadronic cross sections theoretically. The present study explores and compares the three most global PDF sets, viz. CT14, MMHT2014 and NNPDF 4.0 by plotting their corresponding gluon distribution functions as well as the proton structure functions over a wide range of momentum fraction \textit{x} and energy scale \textit{Q}. The plots obtained thus provide significant information about the behaviour of the PDFs.

\section{Graphical analyses and discussion}
\label{section3}
The precise determination of PDFs is important in the LHC age because its uncertainty plays as a limiting factor in the correct QCD theoretical predictions. As PDFs cannot be evaluated using perturbation theory, they must successively be implied from the available experimental data. The comprehension of LHC physics needs proper knowledge of PDFs. The determination of PDFs is done by several international collaborations around the globe. PDFs are needed in various hard-scattering experiments for detailed analyses and interpretation as well as to understand their uncertainties, which are actually important for future experiments.\\

	CT, NNPDF and MMHT are the main successful collaborations associated with the determination of PDFs from all available data, e.g. HERA and LHC. While NNPDF use neural networks for extracting PDFs and Monte-Carlo methods to determine the statistical distribution, the most widely used CT PDFs are based on next-to-leading order (NLO) perturbation theory in the modified minimal subtraction renormalization scheme.\\
	
\begin{figure}[h]
\centering
\includegraphics[scale=0.38]{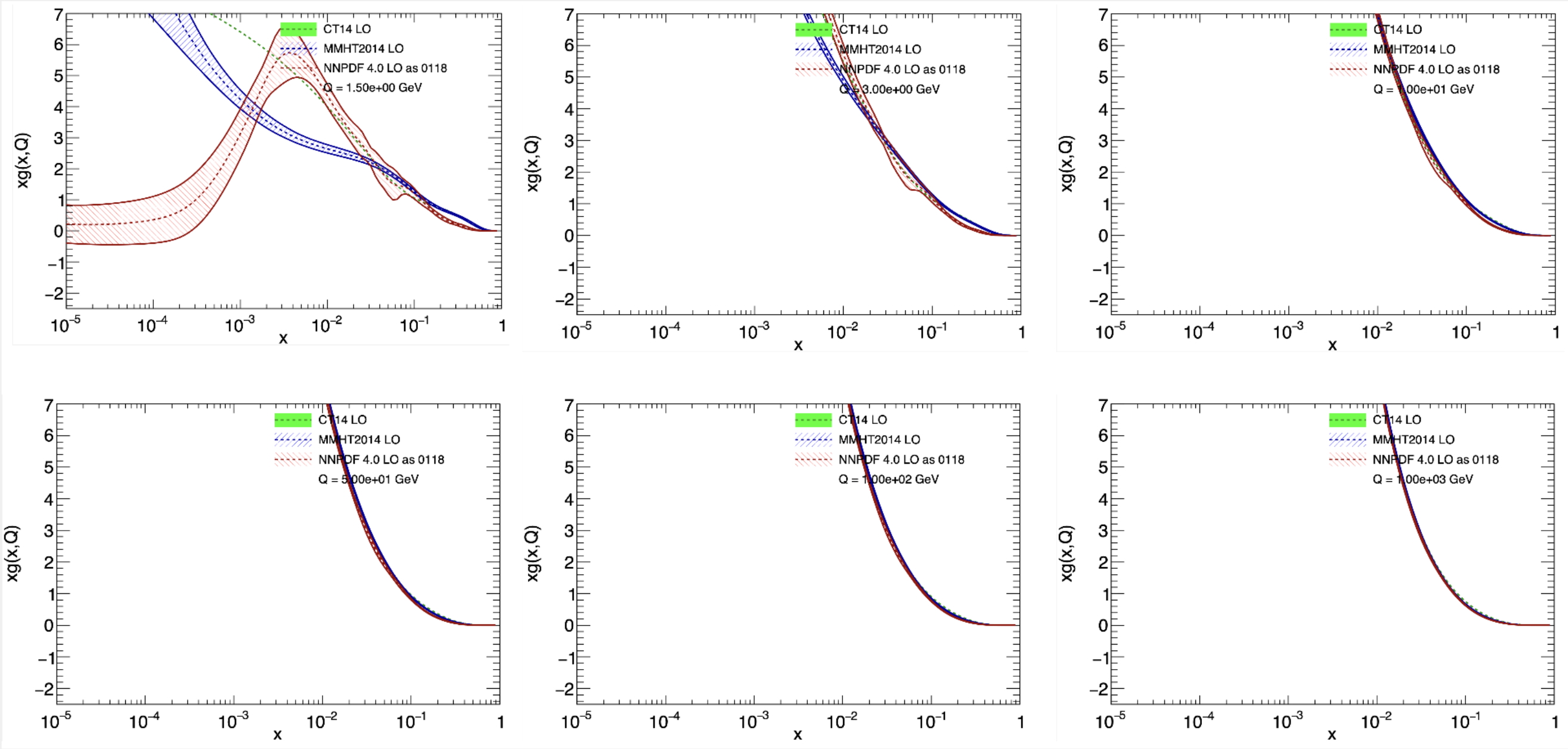}



\caption{Gluon distribution function versus momentum fraction \textit{x} at fixed energy scales \textit{Q} of 1.5, 3, 10, 50, 100 and 1000 GeV.}
\label{Figure1}
\end{figure}

\begin{figure}[h]
\centering
\includegraphics[scale=0.36]{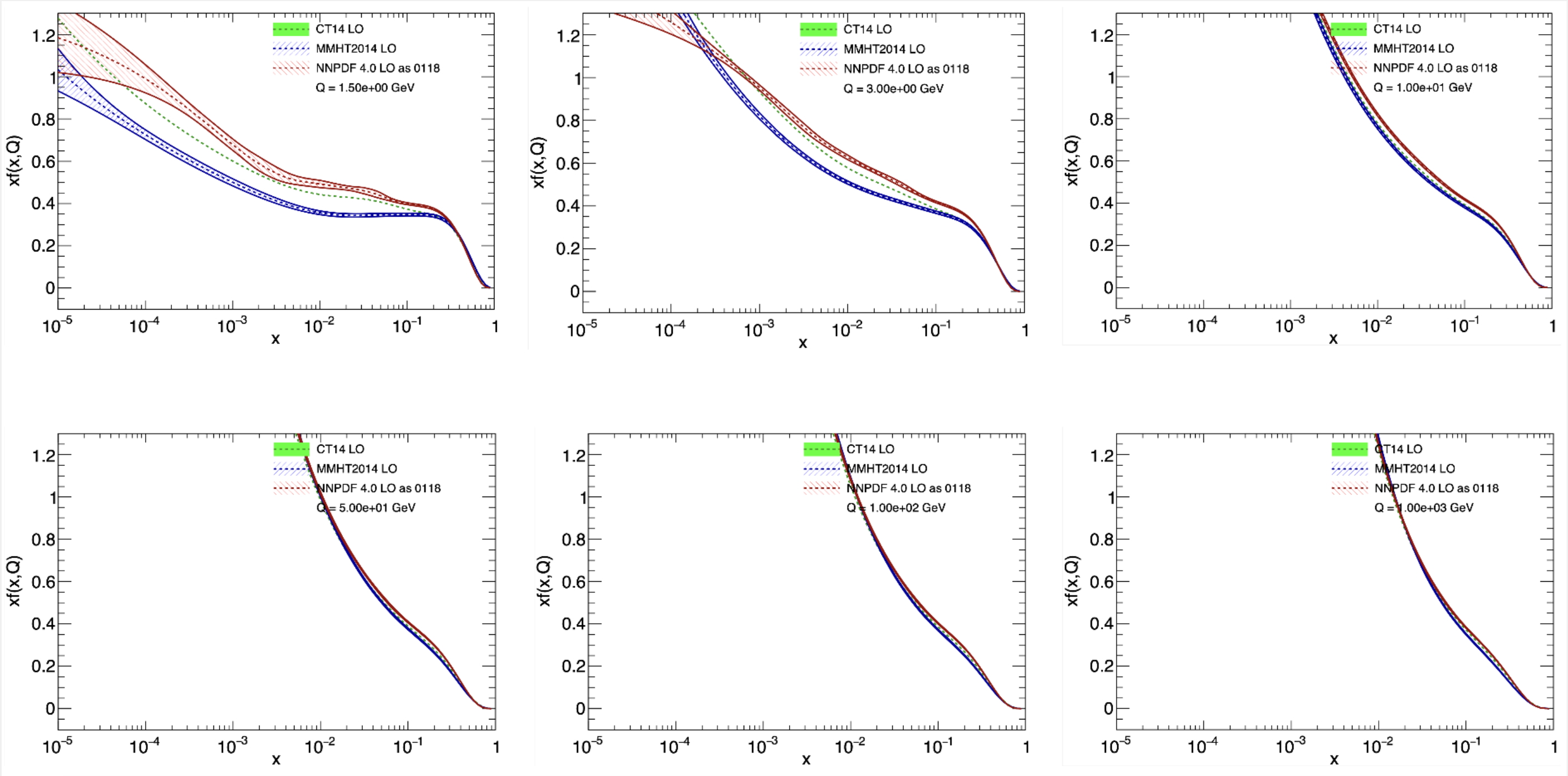}
\caption{Proton structure function versus momentum fraction \textit{x} at fixed energy scales \textit{Q} of 1.5, 3, 10, 50, 100 and 1000 GeV.}
\label{Figure2}
\end{figure}

	Gluon distribution functions have been compared for the three global PDF sets at different values of energy scale \textit{Q} and Bjorken-\textit{x} as shown in Figure \ref{Figure1}. The plots vividly show that gluons are far more abundant at small \textit{x} but decline sharply as \textit{x} increases. Gluons emit either quark-antiquark pairs or extra gluons at large \textit{Q}, and thus result in low parton momenta. It can also be noted that gluons can range negative at small \textit{x}. The behaviour of proton structure functions have also been compared for these PDF sets for different values of \textit{Q} and Bjorken-\textit{x} as shown in Figure \ref{Figure2}. In this case also, the PDFs decline towards large \textit{x} and that they fall sharply at large \textit{Q}. The slight rise of PDFs in the intermediate \textit{x} region might be because of some particular dynamics of higher orders.

\clearpage

\begin{figure}[h]
\centering
\includegraphics[scale=0.31]{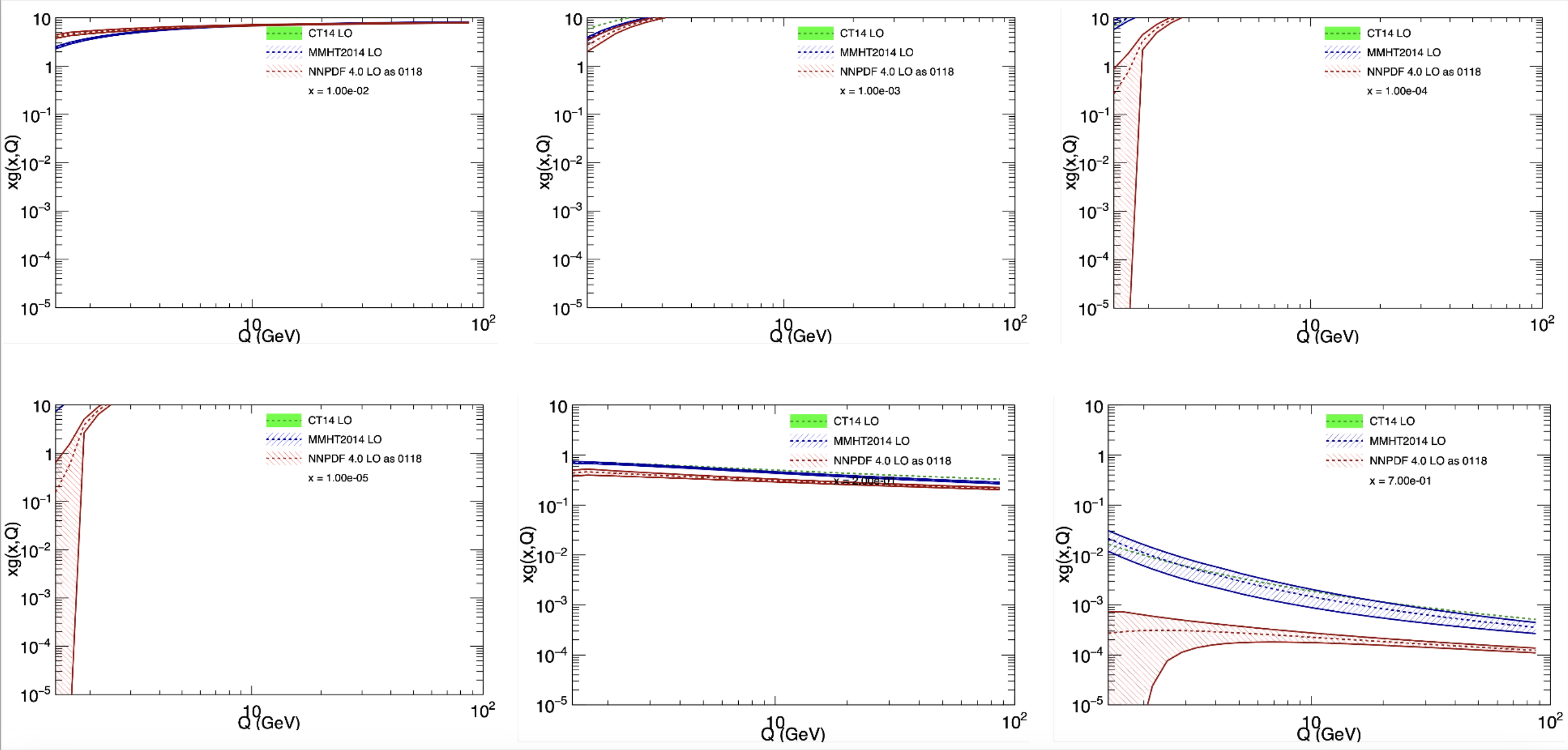}
\caption{Gluon distribution function versus energy scale \textit{Q} at fixed \textit{x} of values $10^{-2}, 10^{-3}, 10^{-4}, 10^{-5}, 0.2$ and $0.7$.}
\label{Figure3}
\end{figure}

\begin{figure}[h]
\centering
\includegraphics[scale=0.31]{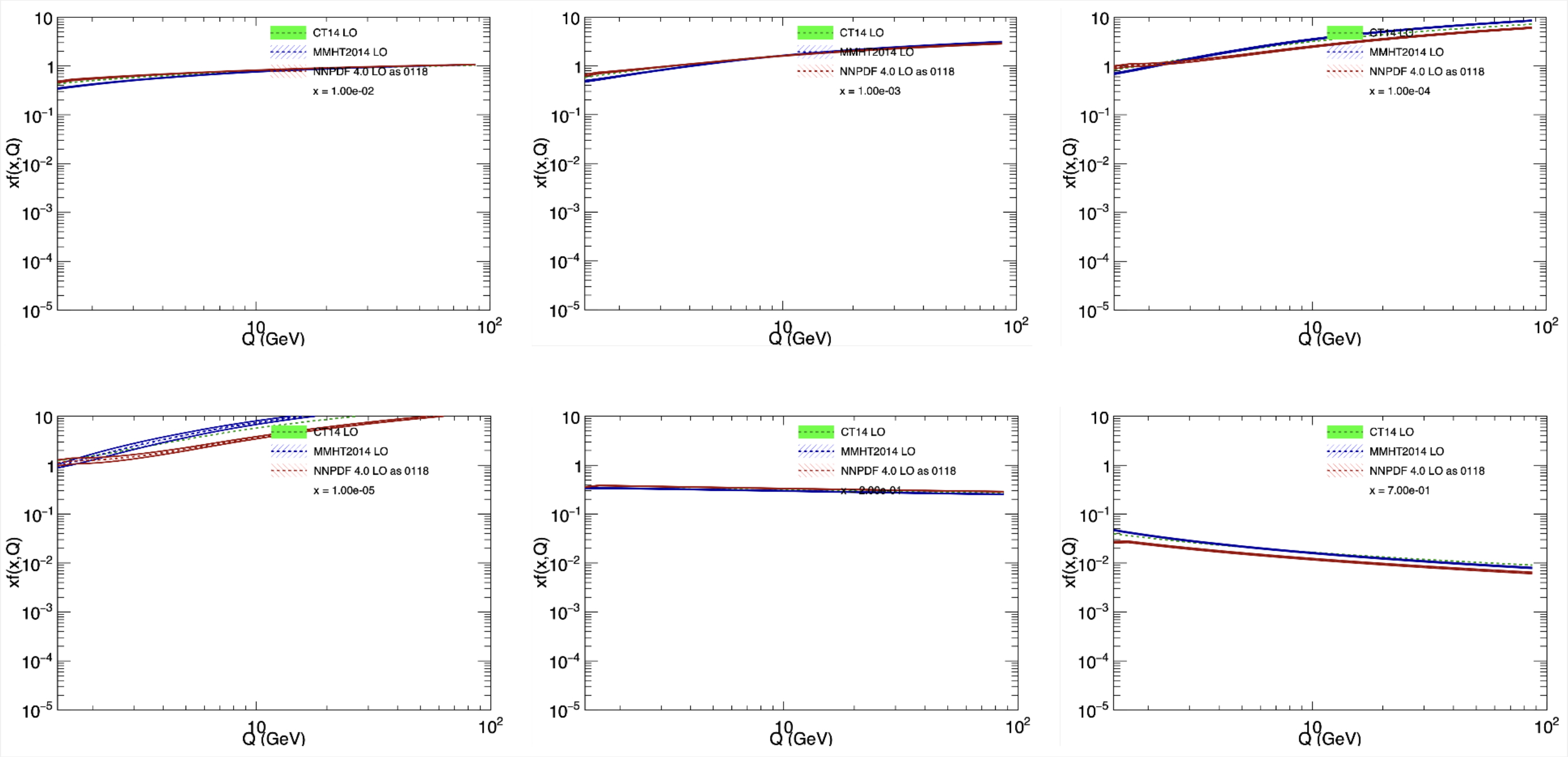}
\caption{Proton structure function versus energy scale \textit{Q} at fixed \textit{x} of values $10^{-2}, 10^{-3}, 10^{-4}, 10^{-5}, 0.2$ and $0.7$.}
\label{Figure4}
\end{figure}

Figure \ref{Figure3} and Figure \ref{Figure4} respectively show the graphical comparison of gluon distribution functions and proton structure functions for the three global PDF sets at fixed \textit{x} and flavour against the energy scale \textit{Q}. While the PDFs rise as \textit{Q} increases at small \textit{x} regime, they fall with \textit{Q} towards large \textit{x}. These patterns are congruent with the theoretical expectations.

\clearpage
In both the figures, it can be observed that the gluon distributions and the proton structure functions fall gradually in the hard-scattering process at large \textit{x} region $(x>0.5)$. The gradual decline of both the functions are less steeper in the valence \textit{x} region $(0.1<x<0.5)$. In the small \textit{x} region $(10^{-4}<x<0.1)$, their behaviours seem to reverse as both the functions rise with \textit{Q}. The rise in the gluon distribution function and proton structure function of the three global PDF sets is much more in the very small \textit{x} region $(x<10^{-4})$.\\

\begin{figure}[h]
\centering
\includegraphics[scale=0.48]{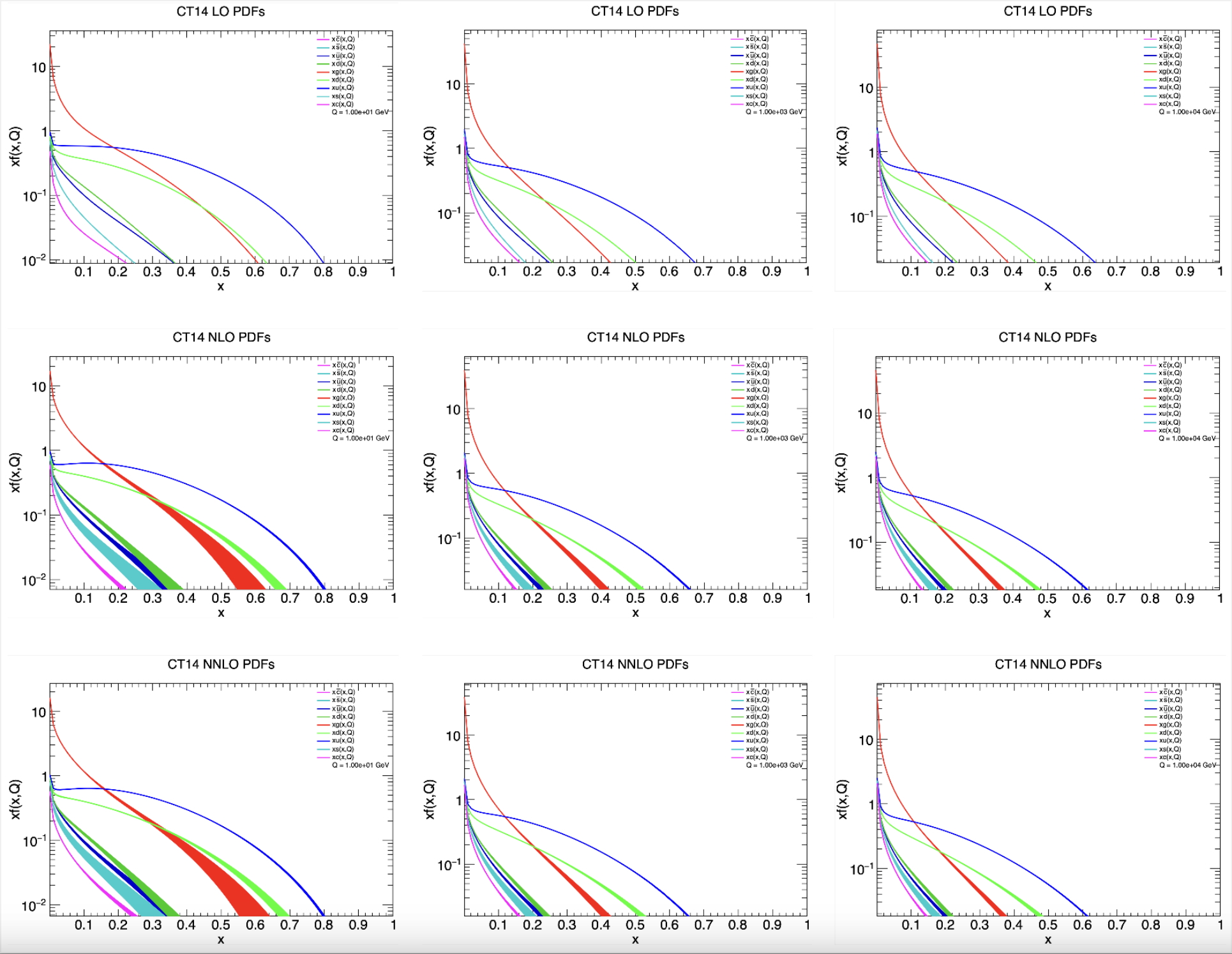}
\caption{CT14 LO, NLO and NNLO PDFs at different values of energy scale \textit{Q}.}
\label{Figure5}
\end{figure}

\begin{figure}[h]
\centering
\includegraphics[scale=0.5]{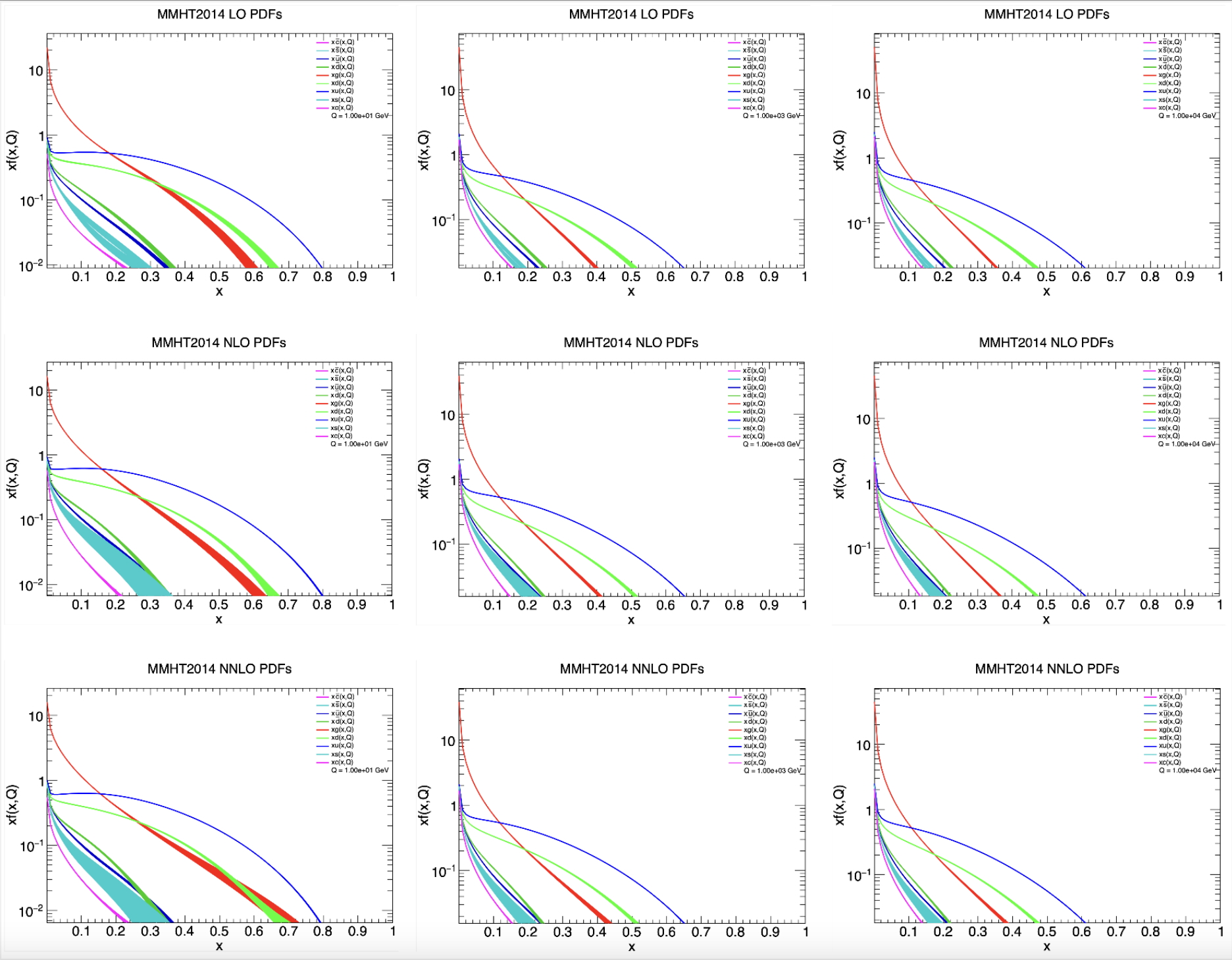}
\caption{MMHT2014 LO, NLO and NNLO PDFs at different values of energy scale \textit{Q}.}
\label{Figure6}
\end{figure}

\begin{figure}[h]
\centering
\includegraphics[scale=0.48]{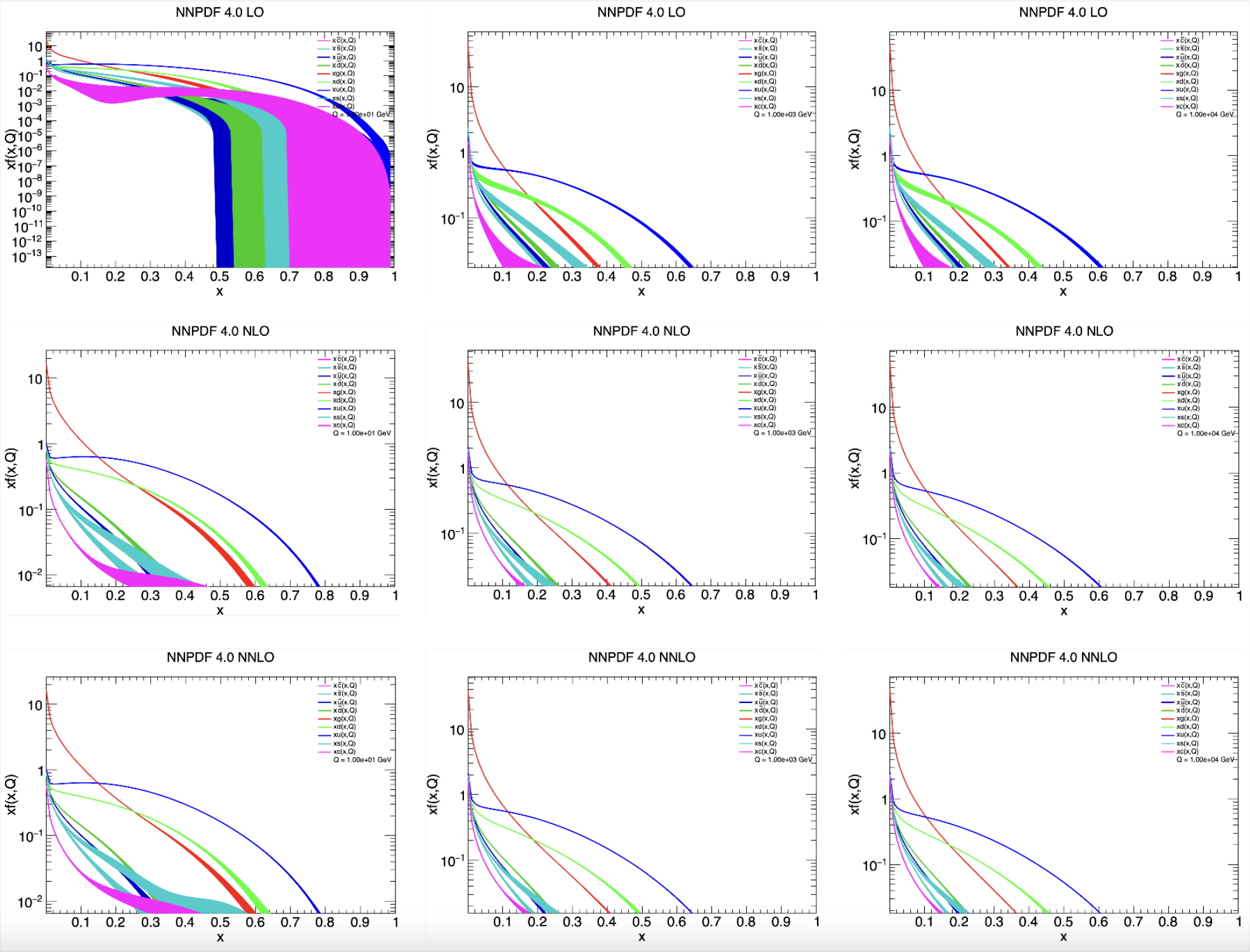}
\caption{NNPDF 4.0 LO, NLO and NNLO PDFs at different values of energy scale \textit{Q}.}
\label{Figure7}
\end{figure}

The CT14, MMHT2014 and NNPDF 4.0 LO, NLO and NNLO PDFs have been graphically evaluated at a few representative values of the energy scale \textit{Q} in Figure \ref{Figure5}, Figure \ref{Figure6} and Figure \ref{Figure7} respectively. The patterns are distinctly understood from the evolution equations. The distributions of \textit{u} and \textit{d} quarks dominate at large \textit{x}, with $u>d$. Gluons are seen to overtop at small \textit{x} because they radiate gluons and their distribution tend to get steeper. Since gluons also create quark-antiquark pairs, there arises a good number of antiquarks at small \textit{x} and their distributions become steeper than the gluons.\\

\clearpage

\clearpage

\section{Conclusion and future outlook}
\label{section4}
The comparative study of global PDFs and proton structure functions is a linchpin in our understanding in QCD. The significance of their comparative study is that by comparing the observed proton structure functions with predictions made from theoretical PDFs, one can validate the underlying principles of QCD at different energy scales. The recursive process of updating PDFs with new experimental data and improved theoretical understanding helps in improving our grasp of proton structure, because it is essential for predicting the cross-sections of several processes in accelerators like the LHC, which are sensitive to the gluon distribution inside protons. These predictions are crucial for designing experiments and for interpreting the collision results, especially in searches for new particles and interactions beyond the Standard Model of particle physics. Specifically, this comparative study of global PDFs and proton structure functions is fundamental for advancing our comprehension of the microscopic world governed by quantum mechanics and for ensuring the accuracy of predictions with regard to particle interactions. The global analysis of PDFs necessitates incorporating data from various experiments around the world and thus helps in creating a more exhaustive model of parton behaviour.\\
	
	PDFs have undoubtedly been an indispensable framework for precision studies and new physics searches at the LHC as well as other experiments involving hadron targets. Since determination of PDFs are crucial for better elucidation of the present and future collider experiments, advanced theoretical approaches and computations of uncertainty analysis need to be developed well. We have discussed a precise and straightforward study on the behaviour of the global PDFs at both small and large Bjorken-\textit{x}.\\
	
	In this paper, graphical analyses and the pertinent discussions of the three global data sets, viz. CT/CTEQ, MMHT and NNPDF, have been done. The behaviour of these PDFs have been thoroughly discussed at both small and large \textit{x} regimes. While gluon distribution function and proton structure function have been plotted against \textit{x} at fixed \textit{Q} in Figure \ref{Figure1} and Figure \ref{Figure2} respectively, they have been plotted against \textit{Q} at fixed \textit{x} in Figure \ref{Figure3} and Figure \ref{Figure4} respectively. The cross section of the distributions falls with the rise in energy scale \textit{Q} due to the fact that the partons lose their momenta by radiating gluons that subsequently split into quark-antiquark pairs. These pairs continue to emit and thus the PDFs increase at small \textit{x}. The LO, NLO and NNLO PDFs of the three global PDF sets have also been explored in Figure \ref{Figure5}, Figure \ref{Figure6} and Figure \ref{Figure7} respectively. This field of research work can be explored further to check the wider applicability of such a PDF tool.
	
\section{Declaration}
No funding has been received for conducting this study. The authors declare no conflicts of interest that are relevant to the content of this manuscript.

\end{document}